\documentclass[aps,prl,reprint,twocolumn,superscriptaddress]{revtex4-1}
\usepackage{amsmath,amssymb}
\usepackage{graphicx}
\usepackage{txfonts}
\usepackage{bm}
\usepackage[cal=boondoxo,calscaled=1.00]{mathalfa}
\usepackage{nicefrac}
\usepackage{color,ulem}
\usepackage{pdfpages}
\usepackage[colorlinks,hyperindex,linkcolor=blue,citecolor=blue,urlcolor=blue]{hyperref}

\newcommand{\bfa}{\mathbf{a}}
\newcommand{\bfC}{\mathbf{C}}
\newcommand{\bfcalC}{{ \bm{\mathcal{C}} }}
\newcommand{\bfe}{\mathbf{e}}
\newcommand{\bfF}{\mathbf{F}}
\newcommand{\bfI}{\mathbf{I}}
\newcommand{\bfJ}{\mathbf{J}}
\newcommand{\bfL}{\mathbf{L}}
\newcommand{\bfm}{\mathbf{m}}
\newcommand{\bfM}{\mathbf{M}}
\newcommand{\bfn}{\mathbf{n}}
\newcommand{\bfP}{\mathbf{P}}
\newcommand{\bfPi}{\mathbf{\Pi}}
\newcommand{\bfq}{\mathbf{q}}
\newcommand{\bfQ}{\mathbf{Q}}
\newcommand{\bfr}{\mathbf{r}}
\newcommand{\bfR}{\mathbf{R}}
\newcommand{\bfS}{\mathbf{S}}
\newcommand{\bfu}{\mathbf{u}}
\newcommand{\bfx}{\mathbf{x}}
\newcommand{\bfxi}{{ \bm{\xiup} }}
\newcommand{\bfY}{\mathbf{Y}}
\newcommand{\itOmega}{\mathit{\Omega}}
\newcommand{\barf}{\bar{f}}
\newcommand{\barg}{\bar{g}}
\newcommand{\barbfm}{\bar{\mathbf{m}}}
\newcommand{\barbfn}{\bar{\mathbf{n}}}

\newcommand{\tila}{\tilde{a}}
\newcommand{\tilb}{\tilde{b}}
\newcommand{\add}{\text{add}}
\newcommand{\BE}{\text{\tiny BE}}
\newcommand{\cri}{\text{cr}}
\newcommand{\Cref}{C_\text{ref}}

\newcommand{\eq}{\text{eq}}
\newcommand{\erf}{\text{erf}}

\newcommand{\Enskog}{\text{Enskog}}
\newcommand{\EOS}{\text{\tiny EOS}}
\newcommand{\lv}{{lv}}
\newcommand{\iso}{\text{iso}}
\newcommand{\INT}{\text{\tiny INT}}
\newcommand{\LBE}{\text{\tiny LBE}}
\newcommand{\mean}{\text{mean}}
\newcommand{\pair}{\text{pair}}
\newcommand{\Ste}{\text{Ste}}
\newcommand{\Tr}{\text{T}}

\newcommand{\dx}{\delta_x}
\newcommand{\narrow}[1]{\thinmuskip=#1 \medmuskip=1.333\thinmuskip \thickmuskip=1.667\thinmuskip}
\newcommand{\figdraft}{false}                   

\begin{document}
\title{Mesoscopic Lattice Boltzmann Modeling of the Liquid-Vapor Phase Transition}
\author{Rongzong Huang}
\email{rongzong.huang@csu.edu.cn}
\affiliation{School of Energy Science and Engineering, Central South University, 410083 Changsha, China}
\author{Huiying Wu}
\email{whysrj@sjtu.edu.cn}
\affiliation{School of Mechanical Engineering, Shanghai Jiao Tong University, 200240 Shanghai, China}
\author{Nikolaus A. Adams}
\email{nikolaus.adams@tum.de}
\affiliation{Institute of Aerodynamics and Fluid Mechanics, Technical University of Munich, 85748 Garching, Germany}
\date{April 22, 2021}

\begin{abstract}
    We develop a mesoscopic lattice Boltzmann model for liquid-vapor phase transition by handling the microscopic molecular interaction. The short-range molecular interaction is incorporated by recovering an equation of state for dense gases, and the long-range molecular interaction is mimicked by introducing a pairwise interaction force. Double distribution functions are employed, with the density distribution function for the mass and momentum conservation laws and an innovative total kinetic energy distribution function for the energy conservation law. The recovered mesomacroscopic governing equations are fully consistent with kinetic theory, and thermodynamic consistency is naturally satisfied.
\end{abstract}

\maketitle

Liquid-vapor phase transition is a widespread phenomenon of great importance in many natural and engineering systems. Because of its multiscale nature and macroscopic complexity \cite{Cheng2014, Onuki2005, Boreyko2009, Cira2015, Nikolayev1999, Nikolayev2006}, thermodynamically consistent modeling of liquid-vapor phase transition with the underlying physics is a long-standing challenge, despite extensive studies. Physically speaking, the phase transition is a natural consequence of the molecular interaction at the microscopic level. Therefore, as a mesoscopic technique that can incorporate the underlying microscopic interaction, the lattice Boltzmann (LB) method is advocated as a promising method for modeling multiphase flows with phase transition \cite{He2002, He1998, Luo1998}.

The theory of the LB method for multiphase flows has been extensively studied since the early 1990s \cite{Gunstensen1991, Shan1993, Swift1995, He1998, Luo1998}. However, most studies are inherently limited to isothermal systems, and the theory of the LB method for liquid-vapor phase transition remains largely unexplored. Recently, some liquid-vapor phase transition problems have been simulated by the LB method \cite{Hazi2009, Gong2012, Li2015, Albernaz2017, Qin2019}, where the popular pseudopotential LB model for isothermal systems is adopted to handle the mass and momentum conservation laws, and a supplementary macroscopic governing equation is employed to handle the energy conservation law. Because of the idea of mimicking the microscopic interaction responsible for multiphase flows, the pseudopotential LB model shows great simplicity in both concept and computation. However, it suffers from thermodynamic inconsistency \cite{He2002}, although the coexistence densities could be numerically tuned close to the thermodynamic results. The supplementary macroscopic energy governing equation is extremely complicated \cite{Onuki2005, Laurila2012} and it is artificially simplified with macroscopic assumptions and approximations in previous works \cite{Hazi2009, Gong2012, Li2015, Albernaz2017, Qin2019}. Both thermodynamic consistency and the underlying physics are sacrificed. Moreover, the simplified energy governing equation cannot be recovered from the mesoscopic level, implying that the computational simplicity is also lost.

In this Letter, we first analyze the kinetic model that combines Enskog theory for dense gases with mean-field theory for long-range molecular interaction. Guided by this kinetic model, we develop a novel mesoscopic LB model for liquid-vapor phase transition by handling the underlying microscopic molecular interaction rather than resorting to any macroscopic assumptions or approximations. The present LB model has a clear physical picture at the microscopic level and thus the conceptual and computational simplicity, and it is also kinetically and thermodynamically consistent.

The microscopic molecular interaction responsible for liquid-vapor phase transition generally consists of a short-range repulsive core and a long-range attractive tail. The short-range molecular interaction can be well modeled by Enskog theory for dense gases, and the long-range molecular interaction can be described by mean-field theory and thus modeled as a point force \cite{Rowlinson1982}. Combining Enskog theory for dense gases with mean-field theory for long-range molecular interaction, the kinetic equation for the density distribution function (DF) $f (\bfx, \bfxi, t)$ can be written as \cite{He2002}
\begin{equation}
\partial_t f + \bfxi \cdot \nabla f + \bfa \cdot \nabla_\bfxi f = \itOmega _\Enskog + \nabla V_\mean \cdot \nabla_\bfxi f ,
\end{equation}
where $\bfxi$ is the molecular velocity, $\bfa$ is the external acceleration, and $V_\mean$ denotes the mean-field approximation of the long-range molecular potential. The Enskog collision operator $\itOmega_\Enskog$ is \cite{Chapman1970}
\begin{equation}
\narrow{1.0mu}
\begin{aligned}
\itOmega_\Enskog = \chi \itOmega_0 - b \rho \chi f^\eq \Big\{ \tfrac25 \left[ 2 \bfcalC \bfcalC : \nabla \bfu + \left( | \bfcalC |^2 - \tfrac52 \right) \nabla \cdot \bfu \right] & \\
+ \bfC \cdot \left[ \nabla \ln (\rho^2 \chi T) + \tfrac35 \left( |\bfcalC|^2 - \tfrac52 \right) \nabla \ln T \right] & \Big\} ,
\end{aligned}
\end{equation}
where $\itOmega_0$ is the usual collision operator for rarefied gases, $\chi$ is the collision probability, $b = 2 \pi d^3 \big{/} (3 m)$ with $d$ and $m$ the molecular diameter and mass, $\bfC = \bfxi - \bfu$, and $\bfcalC = \bfC \big{/}\! \sqrt{2 RT}$. The equilibrium DF $f^\eq$ is
\begin{equation}
f^\eq = \dfrac{\rho}{ (2 \pi RT)^{3/2} } \exp \left( - |\bfcalC|^2 \right) .
\end{equation}
The density $\rho$ and momentum $\rho \bfu$ are calculated as 
\begin{equation}
\rho = \int f d\bfxi , \quad \rho \bfu = \int f \bfxi d\bfxi .
\end{equation}
Based on the density DF, a distinct internal kinetic energy $\rho \epsilon_k^{}$ and total kinetic energy $\rho e_k^{}$ can be well defined: 
\begin{equation}
\rho \epsilon_k^{} = \int f \dfrac{|\bfxi - \bfu|^2}{2} d\bfxi, \quad \rho e_k^{} = \int f \dfrac{|\bfxi|^2}{2} d\bfxi .
\end{equation}
Because of the long-range molecular interaction, the internal potential energy, defined as $\rho \epsilon_p = \frac12 \rho V_\mean$, should be considered. Here, the factor $\tfrac12$ avoids counting each interacting pair twice. Therefore, the usual internal energy and total energy are $\rho \epsilon = \rho \epsilon_k^{} + \rho \epsilon_p$ and $\rho e = \rho e_k^{} + \rho \epsilon_p$. Through the Chapman-Enskog (CE) analysis, the following mesomacroscopic governing equations can be derived:
\begin{subequations}\label{eq.conservation}
    \narrow{1.5mu}
    \begin{gather}
    \partial_t \rho + \nabla \cdot (\rho \bfu) = 0, \\
    \label{eq.conservation.mom}
    \partial_t (\rho \bfu) + \nabla \cdot (\rho \bfu \bfu) = -\nabla p_\BE^{} + \bfF_\mean + \rho \bfa + \nabla \cdot \bfPi, \\
    \label{eq.conservation.tke}
    \partial_t (\rho e_k^{}) + \nabla \cdot (\rho e_k^{} \bfu + p_\BE^{} \bfu) = \bfF_\mean \cdot \bfu + \rho \bfa \cdot \bfu + \nabla \cdot (\bfJ + \bfu \cdot \bfPi) ,
    \end{gather}
\end{subequations}
where $p_\BE^{} = \rho R T (1 + b \rho \chi)$ is the equation of state (EOS) for dense gases recovered by the Enskog collision operator, $\bfF_\mean = -\rho \nabla V_\mean$ is the point force for the long-range molecular interaction, $\bfPi$ is the viscous stress tensor, and $\bfJ$ denotes the energy flux by conduction. Note that Eq.\ (\ref{eq.conservation}) should be viewed as mesomacroscopic rather than macroscopic governing equations because the involved $\bfF_\mean$ and $\rho e_k^{}$ cannot be well defined from the macroscopic viewpoint.

Equation (\ref{eq.conservation.tke}) in terms of $\rho e_k^{}$ is uncommon in previous works. To derive the usual macroscopic energy governing equation, the transport equation for $\rho \epsilon_p$ should be first established. The mean-field approximation of the long-range molecular potential is given as \cite{Rowlinson1982}
\begin{equation}\label{eq.Vmean.def}
V_\mean = \int_{ | \bfx_2 - \bfx | > d} \rho (\bfx_2) V(| \bfx_2 - \bfx |) d \bfx_2,
\end{equation}
where $\bfx$ and $\bfx_2$ are the positions of two interacting molecules, $V( |\bfx_2 - \bfx| )$ is the distance-dependent potential. Performing Taylor series expansion of $\rho (\bfx_2)$ centered at $\bfx$, Eq.\ (\ref{eq.Vmean.def}) can be formulated as
\begin{equation}\label{eq.Vmean}
V_\mean = -2 a \rho - \kappa \nabla \cdot \nabla \rho ,
\end{equation}
where $a = - \tfrac12 \int_{ | \bfr | > d} V(|\bfr|) d\bfr$ and $\kappa = - \tfrac16 \int_{ | \bfr | > d} |\bfr|^2 V(|\bfr|) d\bfr$. Then, the following relation can be derived:
\begin{equation}
\narrow{1.5mu}
\begin{aligned}
\partial_t (\rho \epsilon_p) + \nabla \cdot (\rho \epsilon_p \bfu) + \bfF_\mean \cdot \bfu = \tfrac12 \rho ( \partial_t V_\mean - \bfu \cdot \nabla V_\mean ) & \\
= - \nabla \cdot [ \bfu \cdot (\bfP - p_\BE^{} \bfI) ] 
+ \nabla \bfu  : \left[ -\tfrac{\kappa}{2} \nabla \cdot (\rho \nabla \rho) \bfI + \kappa \nabla (\rho \nabla \rho) \right] & ,
\end{aligned}
\end{equation}
where $\bfI$ is the unit tensor, and $\bfP$ is the pressure tensor defined as $\nabla \cdot \bfP = \nabla p_\BE^{} - \bfF_\mean$ based on Eq.\ (\ref{eq.conservation.mom}). Adding $\rho \epsilon_p$ to $\rho e_k^{}$, Eq.\ (\ref{eq.conservation.tke}) can be rewritten in terms of $\rho e$:
\begin{equation}\label{eq.conservation.te}
\narrow{1.5mu}
\begin{aligned}
\partial_t (\rho e) + \nabla \cdot (\rho e \bfu + \bfu \cdot \bfP) = \rho \bfa \cdot \bfu + \nabla \cdot (\bfJ + \bfu \cdot \bfPi) & \\
+ \nabla \bfu  : \left[ -\tfrac{\kappa}{2} \nabla \cdot (\rho \nabla \rho) \bfI + \kappa \nabla (\rho \nabla \rho) \right] & .
\end{aligned}
\end{equation}
The last term in Eq.\ (\ref{eq.conservation.te}) refers to the work done by surface tension \cite{He2002}. Equations (\ref{eq.conservation.tke}) and (\ref{eq.conservation.te}) are physically equivalent to each other, but Eq.\ (\ref{eq.conservation.tke}) is much simpler than Eq.\ (\ref{eq.conservation.te}). This is because $\rho e_k^{}$, as a moment of the density DF, is more easily calculated than $\rho e$ at the mesoscopic level, although $\rho e$ is extensively involved at the macroscopic level. Inspired by the above analysis, we will develop a mesoscopic LB model to recover Eq.\ (\ref{eq.conservation}) rather than Eq.\ (\ref{eq.conservation.te}), and the key points are recovering a nonideal-gas EOS like $p_\BE^{}$ that corresponds to the short-range molecular interaction and mimicking the long-range molecular interaction. Note that both the short- and long-range molecular interactions should be included in physically modeling liquid-vapor phase transition. Otherwise, the liquid-vapor system will suffer from density collapse or be homogenized. Before proceeding further, some discussion on the kinetic model is useful. With Eq.\ (\ref{eq.Vmean}), the pressure tensor can be calculated as
\begin{equation}\label{eq.P}
\narrow{1.5mu}
\bfP = \left( p_\EOS^{} - \kappa \rho \nabla \cdot \nabla \rho - \tfrac{\kappa}{2} \nabla \rho \cdot \nabla \rho \right) \bfI + \kappa \nabla \rho \nabla \rho ,
\end{equation}
where $p_\EOS^{} = p_\BE^{} - a \rho^2$ is the full EOS. Obviously, the above $\bfP$ is consistent with thermodynamic theory. The internal kinetic energy is $\rho \epsilon_k^{} = \rho c_v T$ according to kinetic theory, and the total kinetic energy satisfies $\rho e_k^{} = \rho \epsilon_k^{} + \tfrac12 \rho |\bfu|^2$, where $c_v$ is the constant-volume specific heat. The latent heat of vaporization is $h_\lv = h_v - h_l = a (\rho_l^{} - \rho_v) + p_s^{} \big( \rho_v^{-1} - \rho_l^{-1} \big)$, where $h_v$ and $h_l$ are the specific enthalpies ($h = \epsilon + p_\EOS^{} / \rho$) of the saturated vapor and liquid, respectively, $\rho_v$ and $\rho_l^{}$ are the saturated vapor and liquid densities, respectively, and $p_s^{}$ is the saturation pressure.

Based on Eq.\ (\ref{eq.conservation}) derived from the kinetic model, we introduce double DFs: the density DF $f_i (\bfx, t)$ for the mass and momentum conservation laws and an innovative, simple yet effective, total kinetic energy DF $g_i (\bfx, t)$ for the energy conservation law. The standard D2Q9 lattice \cite{Qian1992} is considered here for simplicity, and the extension to three dimensions is straightforward. The LB equations for $f_i$ and $g_i$ are given as 
\begin{subequations}
    \narrow{1.5mu}
    \begin{gather}
    \label{eq.streaming}
    \ell_i (\bfx + \bfe_i \delta_t, t + \delta_t) = \bar{\ell}_i (\bfx, t), \\
    \label{eq.collision.m}
    \barbfm = \bfm + \delta_t \bfF_m - \bfS \left( \bfm - \bfm^\eq + \tfrac{\delta_t}{2} \bfF_m \right) + \bfS \bfQ_m, \\
    \label{eq.collision.n}
    \barbfn = \bfn + \delta_t \bfq_m - \bfL \left( \bfn - \bfn^\eq + \tfrac{\delta_t}{2} \bfq_m \right) + c^2 \bfY \left( \tfrac{ \bfm + \barbfm }{2} - \bfm^\eq \right) ,
    \end{gather}
\end{subequations}
where Eq.\ (\ref{eq.streaming}) is the linear streaming process in velocity space with $\ell_i$ denoting $f_i$ or $g_i$ and the overbar denoting the post-collision state, Eqs.\ (\ref{eq.collision.m}) and (\ref{eq.collision.n}) are the local collision processes in moment space computed at position $\bfx$ and time $t$ with the moments $\bfm = \bfM (f_i)^\Tr$ and $\bfn = \bfM (g_i)^\Tr$, and $c = \delta_x / \delta_t$ is the lattice speed. The post-collision DFs are obtained via $(\barf_i)^\Tr = \bfM^{-1} \barbfm$ and $(\barg_i)^\Tr = \bfM^{-1} \barbfn$. Here, $\bfM$ is the orthogonal transformation matrix \cite{Lallemand2000}. A pairwise interaction force is introduced to mimic the long-range molecular interaction, which is given as \cite{Huang2019.STEOS}
\begin{equation}
\bfF_\pair = G^2 \rho (\bfx) \sum\nolimits_i \omega (|\bfe_i \delta_t|^2) \rho (\bfx + \bfe_i \delta_t) \bfe_i \delta_t, 
\end{equation}
where $G^2$ controls the interaction strength, $\omega (\delta_x^2) = \tfrac13$ and $\omega (2\delta_x^2) = \tfrac{1}{12}$ maximize the isotropy degree of $\bfF_\pair$. The density $\rho$, momentum $\rho \bfu$, and total kinetic energy $\rho e_k^{}$ are calculated as
\begin{equation}\label{eq.sumup}
    \narrow{1.5mu}
    \rho = \sum\nolimits_i f_i, \quad \rho \bfu = \sum\nolimits_i \bfe_i f_i + \tfrac{\delta_t}{2} \bfF, \quad \rho e_k^{} = \sum\nolimits_i g_i + \tfrac{\delta_t}{2} q .
\end{equation}
Here, $\bfF = \bfF_\pair + \rho \bfa$ is the total force, and $q = \bfF_\pair \cdot \bfu + \rho \bfa \cdot \bfu$ is the total work done by force. Note that $\tfrac{\delta_t}{2} \bfF$ and $\tfrac{\delta_t}{2} q$ in Eq.\ (\ref{eq.sumup}) are necessary to avoid the discrete lattice effect.

The technical details of the present mesoscopic LB model (including the equilibrium moments $\bfm^\eq$ and $\bfn^\eq$, the collision matrices $\bfS$ and $\bfL$, the discrete force $\bfF_m$, the discrete source $\bfq_m$, etc.) are given in Supplemental Material \cite{SM}. Performing the second- and third-order CE analyses for the above LB model, the mesomacroscopic governing equations from the kinetic model [i.e., Eq.\ (\ref{eq.conservation})] can be recovered once we set 
\begin{equation}\label{eq.analogy.1}
\begin{array}{l}
p_\BE^{} = p_\LBE^{}, \quad \bfF_\mean = \bfF_\pair + \bfR_\iso + \bfR_\add, \\[0.7ex]
        \rho h_k = \rho e_k^{} + p_\BE^{},
\end{array}
\end{equation}
where $p_\LBE^{} = c_s^2 (\rho + \eta)$ is the recovered EOS for dense gases with $\eta$ a built-in variable in $\bfm^\eq$, $\bfR_\iso = \tfrac{1}{12} \delta_x^2 \nabla \cdot \nabla \bfF_\pair$ and $\bfR_\add = - \tfrac{G^2 \delta_x^4}{24} \nabla \cdot [2 \nabla \rho \nabla \rho + (\nabla \rho \cdot \nabla \rho) \bfI]$ are the third-order terms by the third-order discrete lattice effect and by the compensation term $\bfS \bfQ_m$ in Eq.\ (\ref{eq.collision.m}), respectively, and $\rho h_k$ is the \textit{total kinetic enthalpy} in $\bfn^\eq$. The recovered viscous stress tensor and energy flux are given as $\bfPi = \rho \nu [ \nabla \bfu + (\nabla \bfu)^\Tr - (\nabla \cdot \bfu) \bfI] + \rho \varsigma (\nabla \cdot \bfu) \bfI$ and $\bfJ = \lambda \nabla T$, respectively, with the kinematic viscosity $\nu = c_s^2 \delta_t \big( s_p^{-1} - \tfrac12 \big)$, the bulk viscosity $\varsigma = \varpi c_s^2 \delta_t \big( s_e^{-1} - \tfrac12 \big)$, and the heat conductivity $\lambda = \tfrac{4 + 3\gamma_1 + 2\gamma_2}{6} \Cref c^2 \delta_t \big( \sigma_j^{-1} - \tfrac12 \big)$. Here, $\varpi$ and $\gamma_{1,2}$ are model coefficients, $s_{e,p}$ and $\sigma_j$ are relaxation parameters, $\Cref$ is the reference \textit{volumetric heat capacity} \cite{SM}, and $c_s = c \big{/}\! \sqrt{3}$ is the lattice sound speed. Based on Eq.\ (\ref{eq.analogy.1}), the pressure tensor given by Eq.\ (\ref{eq.P}) can be derived, and there have
\begin{equation}
a = \tfrac{G^2 \delta_x^2}{2} , \quad \kappa = \tfrac{G^2 \delta_x^4}{4} .
\end{equation}
Therefore, thermodynamic consistency naturally emerges from our mesoscopic LB model developed in accordance with the kinetic model. Note that there exist some additional cubic terms of velocity in recovering the viscous stress tensor \cite{Dellar2014, Geier2018}, which are ignored with the low Mach number condition and can also be eliminated by trivial modifications \cite{Huang2019.STEOS, Huang2020}. Moreover, the present LB model shows satisfactory numerical stability due to the separate incorporations of the short- and long-range molecular interactions and the introduction of an innovative, simple yet effective, total kinetic energy DF.

In this work, the following full EOS combining the Carnahan-Starling expression for hard spheres \cite{Carnahan1969} with an attractive term is specified:
\begin{equation}\label{eq.cseos}
p_\EOS^{} = K_\EOS \left[ \rho RT \tfrac{1 + \vartheta + \vartheta^2 - \vartheta^3}{ (1 - \vartheta)^3 } - \tila \rho^2 \right], 
\end{equation}
where $\vartheta = \tilb \rho /4$, $\tila = 0.4963880577294099 R^2 T_\cri^2 \big{/} p_\cri^{}$, and $\tilb = 0.1872945669467330 R T_\cri \big{/} p_\cri^{}$. Here, $T_\cri$ and $p_\cri^{}$ are the critical temperature and pressure, respectively. The interaction strength is set to
\begin{equation}\label{eq.G}
G = K_\INT \sqrt{2 K_\EOS \tila \big{/} \delta_x^2} \,,
\end{equation}
and the lattice sound speed is chosen as
\begin{equation}\label{eq.cs}
c_s = K_\INT \left. \sqrt{ \left( \tfrac{\partial p_\EOS^{}}{\partial \rho} \right)_T + 2 K_\EOS \tila \rho } \,\right|\, \raisebox{-1.7ex}{$\scriptstyle \rho \,=\, \rho_l^{}$} . 
\end{equation}
Note that the scaling factors $K_\EOS$ and $K_\INT$ are introduced to adjust the surface tension $\sigma \propto K_\EOS K_\INT$ and interface thickness $W \propto K_\INT$.

\begin{figure}[htbp]
  \centering
  \includegraphics[scale=0.99,draft=\figdraft]{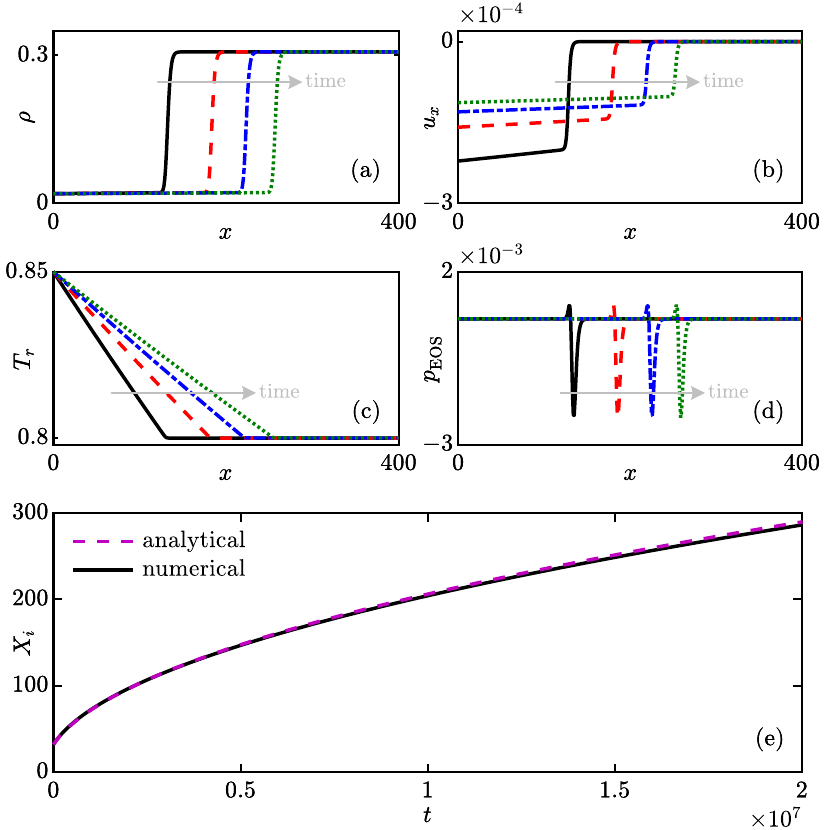}
  \caption{Distributions of (a) density $\rho$, (b) velocity $u_x$, (c) temperature $T_r$, and (d) pressure $p_\EOS^{}$ at time $t = 0.401 \times 10^7$, $0.803 \times 10^7$, $1.204 \times 10^7$, and $1.606 \times 10^7$, and (e) time evolution of the phase interface location $X_i$ for the one-dimensional Stefan problem.}
  \label{Fig.01}
\end{figure}

To test the applicability of our mesoscopic LB model for liquid-vapor phase transition, we perform simulations with $\varpi = 1/6$, $\gamma_1 = -2$, $\gamma_2 = 2$, $\tila = 1$, $\tilb = 4$, $R = 1$, and $\delta_x = 1$. The reduced temperature ($T_r = T / T_\cri$) is set to $T_{r, 0} = 0.8$, and the surface tension $\sigma = 0.01$ and interface thickness $W = 10$, which indicate that $K_\EOS = 0.479820$ and $K_\INT = 2.294922$. The kinematic viscosities and heat conductivities of the liquid and vapor satisfy $\nu_l = \nu_v$ and $\lambda_l = 10 \lambda_v$, respectively. A higher temperature $T_{r,1} = 0.85$, together with the outflow and constant-pressure condition, is applied to drive the phase transition. This boundary condition is treated by the improved nonequilibrium-extrapolation scheme \cite{Huang2019.TLBM}. Meanwhile, eliminating the additional cubic terms of velocity is also plugged into the LB model \cite{Huang2020}. Before simulating liquid-vapor phase transition, an equilibrium droplet in periodic domain is considered. The numerical results of $\sigma$ and $W$, measured by Laplace's law and circular fitting, are $0.0101676$ and $9.961104$, respectively. Such good agreements with the prescribed values validate the present LB model. Subsequently, the one-dimensional Stefan problem is simulated on a $1024 \delta_x \times 4 \delta_x$ domain heated from the left side. Neglecting convection and taking the sharp-interface limit, the analytical location of liquid-vapor phase interface can be obtained \cite{Solomon1966}:
\begin{equation}
X_i(t) = 2 k \sqrt{\alpha_v (t + t_0)} ,
\end{equation}
where $\alpha_v = \lambda_v \big{/} (\rho_v c_v)$, $t_0$ shifts the initial location, and $k$ is the root of the transcendental equation
\begin{equation}
\tfrac{\Ste}{\exp (k^2) \, \erf(k) } = k \sqrt{\pi} ,
\end{equation}
where the Stefan number is defined as $\Ste = \rho_v c_v T_\cri (T_{r,1}  - T_{r,0}) \big{/} (\rho_l^{} h_\lv)$ and set to $\Ste = 0.005$ to ensure that convection can be neglected. The numerical results are shown in Fig.\ \ref{Fig.01}. It can be seen that liquid-vapor phase transition is successfully and accurately captured by the present LB model. The vapor slowly flows to the left with its temperature gradually rising from $T_{r,0}$ to $T_{r,1}$, while the liquid stays at rest with a uniform temperature $T_{r,0}$. Across the phase interface, the density profile can be well maintained, and the pressures in vapor and liquid balance each other (the jumps of $p_\EOS^{}$ within the phase interface come from the nonmonotonic EOS for liquid-vapor fluids). Moreover, the location of phase interface agrees very well with the analytical result, which suggests that the latent heat of vaporization in the mesoscopic LB model is naturally consistent with thermodynamic theory.

\begin{figure}[htbp]
  \centering
  \includegraphics[scale=1,draft=\figdraft]{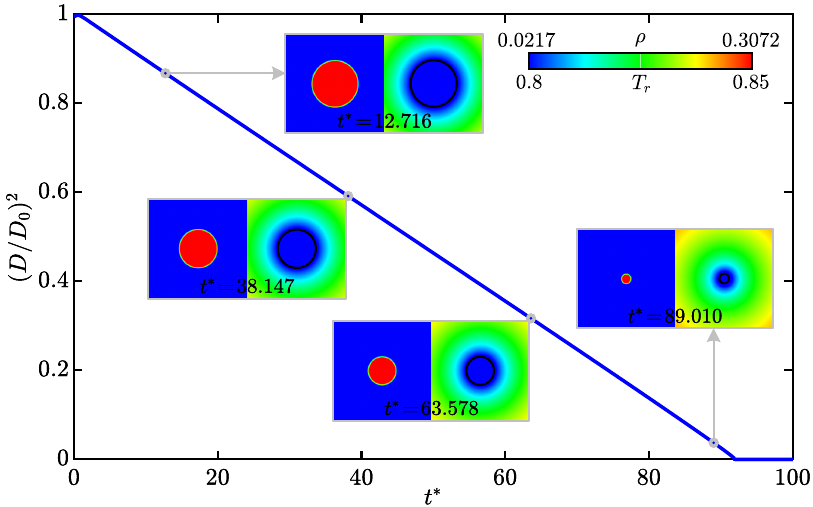}
  \caption{Evolution of the square of the normalized droplet diameter $(D / D_0)^2$ with the normalized time $t^\ast = \alpha_v t \big{/} D_0^2$ in the droplet evaporation process. Snapshots of the local density (left) and temperature (right) fields are also shown at time $t^\ast = 12.716$, $38.147$, $63.578$, and $89.010$, where the solid line in temperature field denotes the liquid-vapor phase interface.}
  \label{Fig.02}
\end{figure}

\begin{figure}[htbp]
  \centering
  \includegraphics[scale=1,draft=\figdraft]{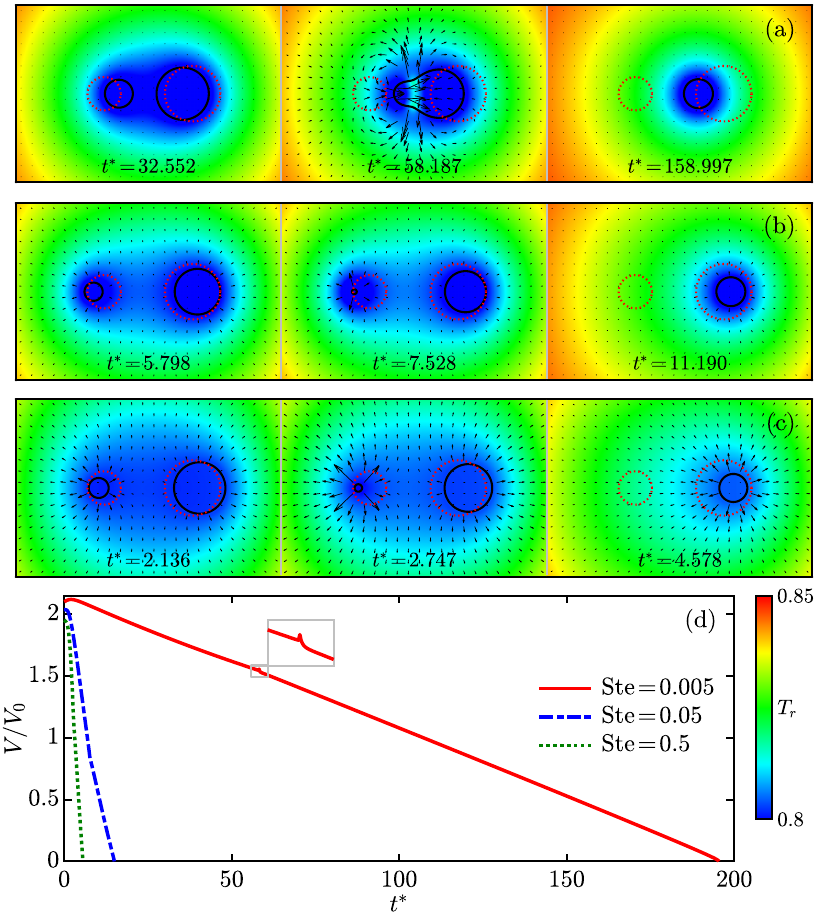}
  \caption{Snapshots of the local temperature and velocity fields at different normalized times for (a) $\Ste = 0.005$, (b) $\Ste = 0.05$, and (c) $\Ste = 0.5$, and (d) evolutions of the normalized total volume of the droplets $V / V_0$ with the normalized time $t^\ast = \alpha_v t \big{/} D_0^2$ in the evaporation process of a large-small droplet pair. The solid and dotted lines in temperature field denote the liquid-vapor phase interface and its initial location, respectively.}
  \label{Fig.03}
\end{figure}

A liquid droplet with diameter $D_0 = 256 \delta_x$ is then simulated on a $1024 \delta_x \times 1024 \delta_x$ domain heated from all the four sides. The Stefan number is set to $\Ste = 0.005$ and thus convection in the evaporation is quite weak. Figure \ref{Fig.02} shows the time evolution of the square of the normalized diameter $(D/D_0)^2$, together with four snapshots of the local density and temperature fields. Here, the time is normalized as $t^\ast = \alpha_v t /D_0^2$. The well-known $D^2$ law \cite{Godsave1953, Safari2013} can be perfectly observed during the entire droplet lifetime, and both the interface thickness and droplet temperature can be well maintained at the prescribed values. As a further application, the evaporation of a large-small droplet pair is simulated with $\Ste = 0.005$, $0.05$, and $0.5$, respectively. Initially, the diameters of the two droplets are $160 \dx$ and $96 \dx$, respectively, and the distance between the droplet centers is $256 \delta_x$. Figure \ref{Fig.03} shows the snapshots of the local temperature and velocity fields, and the time evolution of the normalized volume $V / V_0$. Here, $V$ is the total volume of the droplets, $V_0 = \pi D_0^2 \big{/} 4$, and $D_0 = 128 \dx$. For $\Ste = 0.005$, the evaporation is quite slow, and the two droplets attract each other and coalesce into a single one. This attraction-coalescence behavior is due to the nonuniform evaporation rate along droplet surface, which is induced by the other droplet and will result in an imbalanced vapor recoil force \cite{Nikolayev2006}. Such unusual behavior of evaporating droplets under slow evaporation condition is consistent with the recent experimental and theoretical results \cite{Wen2019, Man2017}. Interestingly, the local temperature slightly rises [see the middle panel in Fig.\ \hyperref[Fig.03]{\ref*{Fig.03}(a)}] and the normalized volume slightly increases [see the ``kink'' in Fig.\ \hyperref[Fig.03]{\ref*{Fig.03}(d)}] when the coalescence occurs, which can be explained as follows: At the neck formed by coalescence, the phase interface changes from convex to concave, and the local saturated vapor pressure will decrease according to the Kelvin equation in thermodynamic theory \cite{Rowlinson1982}. Therefore, the vapor nearby the neck becomes supersaturated and then condenses into liquid, resulting in the release of latent heat and also the increase of droplet volume. Here, it is noteworthy that the above condensation at the neck between two merging droplets is a kind of capillary condensation in thermodynamic theory \cite{Fisher1981, Yang2020}. For $\Ste = 0.05$ and $0.5$, evaporation becomes much faster and convection is very strong. The two droplets repulse each other rather than attract, and the droplet lifetime is much shorter than that for $\Ste = 0.005$. As seen in Fig.\ \hyperref[Fig.03]{\ref*{Fig.03}(b)} and \hyperref[Fig.03]{\ref*{Fig.03}(c)}, the vapor outflows originating from the droplet surfaces impact each other in the middle region between the two droplets, and thus the pressure in this region obviously increases, which then pushes the two droplets away from each other against the imbalanced vapor recoil force.

In summary, we have developed a novel mesoscopic LB model for liquid-vapor phase transition, where the short- and long-range molecular interactions are incorporated by recovering an EOS for dense gases and introducing a pairwise interaction force, respectively, and an innovative, simple yet effective, total kinetic energy DF is proposed for the energy conservation law. The same mesomacroscopic governing equations as the kinetic model can be recovered, and thus thermodynamic consistency is naturally satisfied. Because of the successful modeling of the underlying microscopic molecular interaction, the present mesoscopic LB model does not rely on any macroscopic assumptions or approximations and has the potential to provide reliable physical insights into the liquid-vapor phase transition processes.

\begin{acknowledgments}
R.H.\ acknowledges the support by the Alexander von Humboldt Foundation, Germany. This work was supported by the National Natural Science Foundation of China through Grants No.\ 52006244 and No.\ 51820105009.
\end{acknowledgments}


\begin{thebibliography}{38}%
\makeatletter
\providecommand \@ifxundefined [1]{%
 \@ifx{#1\undefined}
}%
\providecommand \@ifnum [1]{%
 \ifnum #1\expandafter \@firstoftwo
 \else \expandafter \@secondoftwo
 \fi
}%
\providecommand \@ifx [1]{%
 \ifx #1\expandafter \@firstoftwo
 \else \expandafter \@secondoftwo
 \fi
}%
\providecommand \natexlab [1]{#1}%
\providecommand \enquote  [1]{``#1''}%
\providecommand \bibnamefont  [1]{#1}%
\providecommand \bibfnamefont [1]{#1}%
\providecommand \citenamefont [1]{#1}%
\providecommand \href@noop [0]{\@secondoftwo}%
\providecommand \href [0]{\begingroup \@sanitize@url \@href}%
\providecommand \@href[1]{\@@startlink{#1}\@@href}%
\providecommand \@@href[1]{\endgroup#1\@@endlink}%
\providecommand \@sanitize@url [0]{\catcode `\\12\catcode `\$12\catcode
  `\&12\catcode `\#12\catcode `\^12\catcode `\_12\catcode `\%12\relax}%
\providecommand \@@startlink[1]{}%
\providecommand \@@endlink[0]{}%
\providecommand \url  [0]{\begingroup\@sanitize@url \@url }%
\providecommand \@url [1]{\endgroup\@href {#1}{\urlprefix }}%
\providecommand \urlprefix  [0]{URL }%
\providecommand \Eprint [0]{\href }%
\providecommand \doibase [0]{http://dx.doi.org/}%
\providecommand \selectlanguage [0]{\@gobble}%
\providecommand \bibinfo  [0]{\@secondoftwo}%
\providecommand \bibfield  [0]{\@secondoftwo}%
\providecommand \translation [1]{[#1]}%
\providecommand \BibitemOpen [0]{}%
\providecommand \bibitemStop [0]{}%
\providecommand \bibitemNoStop [0]{.\EOS\space}%
\providecommand \EOS [0]{\spacefactor3000\relax}%
\providecommand \BibitemShut  [1]{\csname bibitem#1\endcsname}%
\let\auto@bib@innerbib\@empty
\bibitem [{\citenamefont {Cheng}\ \emph {et~al.}(2014)\citenamefont {Cheng},
  \citenamefont {Quan}, \citenamefont {Gong}, \citenamefont {Liu},\ and\
  \citenamefont {Yang}}]{Cheng2014}%
  \BibitemOpen
  \bibfield  {author} {\bibinfo {author} {\bibfnamefont {P.}~\bibnamefont
  {Cheng}}, \bibinfo {author} {\bibfnamefont {X.}~\bibnamefont {Quan}},
  \bibinfo {author} {\bibfnamefont {S.}~\bibnamefont {Gong}}, \bibinfo {author}
  {\bibfnamefont {X.}~\bibnamefont {Liu}}, \ and\ \bibinfo {author}
  {\bibfnamefont {L.}~\bibnamefont {Yang}},\ }\href@noop {} {\bibfield
  {journal} {\bibinfo  {journal} {Adv. Heat Transf.}\ }\textbf {\bibinfo
  {volume} {46}},\ \bibinfo {pages} {187} (\bibinfo {year} {2014})}\BibitemShut
  {NoStop}%
\bibitem [{\citenamefont {Onuki}(2005)}]{Onuki2005}%
  \BibitemOpen
  \bibfield  {author} {\bibinfo {author} {\bibfnamefont {A.}~\bibnamefont
  {Onuki}},\ }\href@noop {} {\bibfield  {journal} {\bibinfo  {journal} {Phys.
  Rev. Lett.}\ }\textbf {\bibinfo {volume} {94}},\ \bibinfo {pages} {054501}
  (\bibinfo {year} {2005})}\BibitemShut {NoStop}%
\bibitem [{\citenamefont {Boreyko}\ and\ \citenamefont
  {Chen}(2009)}]{Boreyko2009}%
  \BibitemOpen
  \bibfield  {author} {\bibinfo {author} {\bibfnamefont {J.~B.}\ \bibnamefont
  {Boreyko}}\ and\ \bibinfo {author} {\bibfnamefont {C.-H.}\ \bibnamefont
  {Chen}},\ }\href@noop {} {\bibfield  {journal} {\bibinfo  {journal} {Phys.
  Rev. Lett.}\ }\textbf {\bibinfo {volume} {103}},\ \bibinfo {pages} {184501}
  (\bibinfo {year} {2009})}\BibitemShut {NoStop}%
\bibitem [{\citenamefont {Cira}\ \emph {et~al.}(2015)\citenamefont {Cira},
  \citenamefont {Benusiglio},\ and\ \citenamefont {Prakash}}]{Cira2015}%
  \BibitemOpen
  \bibfield  {author} {\bibinfo {author} {\bibfnamefont {N.~J.}\ \bibnamefont
  {Cira}}, \bibinfo {author} {\bibfnamefont {A.}~\bibnamefont {Benusiglio}}, \
  and\ \bibinfo {author} {\bibfnamefont {M.}~\bibnamefont {Prakash}},\
  }\href@noop {} {\bibfield  {journal} {\bibinfo  {journal} {Nature}\ }\textbf
  {\bibinfo {volume} {519}},\ \bibinfo {pages} {446} (\bibinfo {year}
  {2015})}\BibitemShut {NoStop}%
\bibitem [{\citenamefont {Nikolayev}\ and\ \citenamefont
  {Beysens}(1999)}]{Nikolayev1999}%
  \BibitemOpen
  \bibfield  {author} {\bibinfo {author} {\bibfnamefont {V.~S.}\ \bibnamefont
  {Nikolayev}}\ and\ \bibinfo {author} {\bibfnamefont {D.~A.}\ \bibnamefont
  {Beysens}},\ }\href@noop {} {\bibfield  {journal} {\bibinfo  {journal}
  {Europhys. Lett.}\ }\textbf {\bibinfo {volume} {47}},\ \bibinfo {pages} {345}
  (\bibinfo {year} {1999})}\BibitemShut {NoStop}%
\bibitem [{\citenamefont {Nikolayev}\ \emph {et~al.}(2006)\citenamefont
  {Nikolayev}, \citenamefont {Chatain}, \citenamefont {Garrabos},\ and\
  \citenamefont {Beysens}}]{Nikolayev2006}%
  \BibitemOpen
  \bibfield  {author} {\bibinfo {author} {\bibfnamefont {V.~S.}\ \bibnamefont
  {Nikolayev}}, \bibinfo {author} {\bibfnamefont {D.}~\bibnamefont {Chatain}},
  \bibinfo {author} {\bibfnamefont {Y.}~\bibnamefont {Garrabos}}, \ and\
  \bibinfo {author} {\bibfnamefont {D.}~\bibnamefont {Beysens}},\ }\href@noop
  {} {\bibfield  {journal} {\bibinfo  {journal} {Phys. Rev. Lett.}\ }\textbf
  {\bibinfo {volume} {97}},\ \bibinfo {pages} {184503} (\bibinfo {year}
  {2006})}\BibitemShut {NoStop}%
\bibitem [{\citenamefont {He}\ and\ \citenamefont {Doolen}(2002)}]{He2002}%
  \BibitemOpen
  \bibfield  {author} {\bibinfo {author} {\bibfnamefont {X.}~\bibnamefont
  {He}}\ and\ \bibinfo {author} {\bibfnamefont {G.~D.}\ \bibnamefont
  {Doolen}},\ }\href@noop {} {\bibfield  {journal} {\bibinfo  {journal} {J.
  Stat. Phys.}\ }\textbf {\bibinfo {volume} {107}},\ \bibinfo {pages} {309}
  (\bibinfo {year} {2002})}\BibitemShut {NoStop}%
\bibitem [{\citenamefont {He}\ \emph {et~al.}(1998)\citenamefont {He},
  \citenamefont {Shan},\ and\ \citenamefont {Doolen}}]{He1998}%
  \BibitemOpen
  \bibfield  {author} {\bibinfo {author} {\bibfnamefont {X.}~\bibnamefont
  {He}}, \bibinfo {author} {\bibfnamefont {X.}~\bibnamefont {Shan}}, \ and\
  \bibinfo {author} {\bibfnamefont {G.~D.}\ \bibnamefont {Doolen}},\
  }\href@noop {} {\bibfield  {journal} {\bibinfo  {journal} {Phys. Rev. E}\
  }\textbf {\bibinfo {volume} {57}},\ \bibinfo {pages} {R13} (\bibinfo {year}
  {1998})}\BibitemShut {NoStop}%
\bibitem [{\citenamefont {Luo}(1998)}]{Luo1998}%
  \BibitemOpen
  \bibfield  {author} {\bibinfo {author} {\bibfnamefont {L.-S.}\ \bibnamefont
  {Luo}},\ }\href@noop {} {\bibfield  {journal} {\bibinfo  {journal} {Phys.
  Rev. Lett.}\ }\textbf {\bibinfo {volume} {81}},\ \bibinfo {pages} {1618}
  (\bibinfo {year} {1998})}\BibitemShut {NoStop}%
\bibitem [{\citenamefont {Gunstensen}\ \emph {et~al.}(1991)\citenamefont
  {Gunstensen}, \citenamefont {Rothman}, \citenamefont {Zaleski},\ and\
  \citenamefont {Zanetti}}]{Gunstensen1991}%
  \BibitemOpen
  \bibfield  {author} {\bibinfo {author} {\bibfnamefont {A.~K.}\ \bibnamefont
  {Gunstensen}}, \bibinfo {author} {\bibfnamefont {D.~H.}\ \bibnamefont
  {Rothman}}, \bibinfo {author} {\bibfnamefont {S.}~\bibnamefont {Zaleski}}, \
  and\ \bibinfo {author} {\bibfnamefont {G.}~\bibnamefont {Zanetti}},\
  }\href@noop {} {\bibfield  {journal} {\bibinfo  {journal} {Phys. Rev. A}\
  }\textbf {\bibinfo {volume} {43}},\ \bibinfo {pages} {4320} (\bibinfo {year}
  {1991})}\BibitemShut {NoStop}%
\bibitem [{\citenamefont {Shan}\ and\ \citenamefont {Chen}(1993)}]{Shan1993}%
  \BibitemOpen
  \bibfield  {author} {\bibinfo {author} {\bibfnamefont {X.}~\bibnamefont
  {Shan}}\ and\ \bibinfo {author} {\bibfnamefont {H.}~\bibnamefont {Chen}},\
  }\href@noop {} {\bibfield  {journal} {\bibinfo  {journal} {Phys. Rev. E}\
  }\textbf {\bibinfo {volume} {47}},\ \bibinfo {pages} {1815} (\bibinfo {year}
  {1993})}\BibitemShut {NoStop}%
\bibitem [{\citenamefont {Swift}\ \emph {et~al.}(1995)\citenamefont {Swift},
  \citenamefont {Osborn},\ and\ \citenamefont {Yeomans}}]{Swift1995}%
  \BibitemOpen
  \bibfield  {author} {\bibinfo {author} {\bibfnamefont {M.~R.}\ \bibnamefont
  {Swift}}, \bibinfo {author} {\bibfnamefont {W.~R.}\ \bibnamefont {Osborn}}, \
  and\ \bibinfo {author} {\bibfnamefont {J.~M.}\ \bibnamefont {Yeomans}},\
  }\href@noop {} {\bibfield  {journal} {\bibinfo  {journal} {Phys. Rev. Lett.}\
  }\textbf {\bibinfo {volume} {75}},\ \bibinfo {pages} {830} (\bibinfo {year}
  {1995})}\BibitemShut {NoStop}%
\bibitem [{\citenamefont {Hazi}\ and\ \citenamefont {Markus}(2009)}]{Hazi2009}%
  \BibitemOpen
  \bibfield  {author} {\bibinfo {author} {\bibfnamefont {G.}~\bibnamefont
  {Hazi}}\ and\ \bibinfo {author} {\bibfnamefont {A.}~\bibnamefont {Markus}},\
  }\href@noop {} {\bibfield  {journal} {\bibinfo  {journal} {Int. J. Heat Mass
  Tran.}\ }\textbf {\bibinfo {volume} {52}},\ \bibinfo {pages} {1472} (\bibinfo
  {year} {2009})}\BibitemShut {NoStop}%
\bibitem [{\citenamefont {Gong}\ and\ \citenamefont {Cheng}(2012)}]{Gong2012}%
  \BibitemOpen
  \bibfield  {author} {\bibinfo {author} {\bibfnamefont {S.}~\bibnamefont
  {Gong}}\ and\ \bibinfo {author} {\bibfnamefont {P.}~\bibnamefont {Cheng}},\
  }\href@noop {} {\bibfield  {journal} {\bibinfo  {journal} {Int. J. Heat Mass
  Tran.}\ }\textbf {\bibinfo {volume} {55}},\ \bibinfo {pages} {4923} (\bibinfo
  {year} {2012})}\BibitemShut {NoStop}%
\bibitem [{\citenamefont {Li}\ \emph {et~al.}(2015)\citenamefont {Li},
  \citenamefont {Kang}, \citenamefont {Francois}, \citenamefont {He},\ and\
  \citenamefont {Luo}}]{Li2015}%
  \BibitemOpen
  \bibfield  {author} {\bibinfo {author} {\bibfnamefont {Q.}~\bibnamefont
  {Li}}, \bibinfo {author} {\bibfnamefont {Q.~J.}\ \bibnamefont {Kang}},
  \bibinfo {author} {\bibfnamefont {M.~M.}\ \bibnamefont {Francois}}, \bibinfo
  {author} {\bibfnamefont {Y.~L.}\ \bibnamefont {He}}, \ and\ \bibinfo {author}
  {\bibfnamefont {K.~H.}\ \bibnamefont {Luo}},\ }\href@noop {} {\bibfield
  {journal} {\bibinfo  {journal} {Int. J. Heat Mass Tran.}\ }\textbf {\bibinfo
  {volume} {85}},\ \bibinfo {pages} {787} (\bibinfo {year} {2015})}\BibitemShut
  {NoStop}%
\bibitem [{\citenamefont {Albernaz}\ \emph {et~al.}(2017)\citenamefont
  {Albernaz}, \citenamefont {Do-Quang}, \citenamefont {Hermanson},\ and\
  \citenamefont {Amberg}}]{Albernaz2017}%
  \BibitemOpen
  \bibfield  {author} {\bibinfo {author} {\bibfnamefont {D.~L.}\ \bibnamefont
  {Albernaz}}, \bibinfo {author} {\bibfnamefont {M.}~\bibnamefont {Do-Quang}},
  \bibinfo {author} {\bibfnamefont {J.~C.}\ \bibnamefont {Hermanson}}, \ and\
  \bibinfo {author} {\bibfnamefont {G.}~\bibnamefont {Amberg}},\ }\href@noop {}
  {\bibfield  {journal} {\bibinfo  {journal} {J. Fluid Mech.}\ }\textbf
  {\bibinfo {volume} {820}},\ \bibinfo {pages} {61} (\bibinfo {year}
  {2017})}\BibitemShut {NoStop}%
\bibitem [{\citenamefont {Qin}\ \emph {et~al.}(2019)\citenamefont {Qin},
  \citenamefont {Del~Carro}, \citenamefont {Mazloomi~Moqaddam}, \citenamefont
  {Kang}, \citenamefont {Brunschwiler}, \citenamefont {Derome},\ and\
  \citenamefont {Carmeliet}}]{Qin2019}%
  \BibitemOpen
  \bibfield  {author} {\bibinfo {author} {\bibfnamefont {F.}~\bibnamefont
  {Qin}}, \bibinfo {author} {\bibfnamefont {L.}~\bibnamefont {Del~Carro}},
  \bibinfo {author} {\bibfnamefont {A.}~\bibnamefont {Mazloomi~Moqaddam}},
  \bibinfo {author} {\bibfnamefont {Q.}~\bibnamefont {Kang}}, \bibinfo {author}
  {\bibfnamefont {T.}~\bibnamefont {Brunschwiler}}, \bibinfo {author}
  {\bibfnamefont {D.}~\bibnamefont {Derome}}, \ and\ \bibinfo {author}
  {\bibfnamefont {J.}~\bibnamefont {Carmeliet}},\ }\href@noop {} {\bibfield
  {journal} {\bibinfo  {journal} {J. Fluid Mech.}\ }\textbf {\bibinfo {volume}
  {866}},\ \bibinfo {pages} {33} (\bibinfo {year} {2019})}\BibitemShut
  {NoStop}%
\bibitem [{\citenamefont {Laurila}\ \emph {et~al.}(2012)\citenamefont
  {Laurila}, \citenamefont {Carlson}, \citenamefont {Do-Quang}, \citenamefont
  {Ala-Nissila},\ and\ \citenamefont {Amberg}}]{Laurila2012}%
  \BibitemOpen
  \bibfield  {author} {\bibinfo {author} {\bibfnamefont {T.}~\bibnamefont
  {Laurila}}, \bibinfo {author} {\bibfnamefont {A.}~\bibnamefont {Carlson}},
  \bibinfo {author} {\bibfnamefont {M.}~\bibnamefont {Do-Quang}}, \bibinfo
  {author} {\bibfnamefont {T.}~\bibnamefont {Ala-Nissila}}, \ and\ \bibinfo
  {author} {\bibfnamefont {G.}~\bibnamefont {Amberg}},\ }\href@noop {}
  {\bibfield  {journal} {\bibinfo  {journal} {Phys. Rev. E}\ }\textbf {\bibinfo
  {volume} {85}},\ \bibinfo {pages} {026320} (\bibinfo {year}
  {2012})}\BibitemShut {NoStop}%
\bibitem [{\citenamefont {Rowlinson}\ and\ \citenamefont
  {Widom}(1982)}]{Rowlinson1982}%
  \BibitemOpen
  \bibfield  {author} {\bibinfo {author} {\bibfnamefont {J.~S.}\ \bibnamefont
  {Rowlinson}}\ and\ \bibinfo {author} {\bibfnamefont {B.}~\bibnamefont
  {Widom}},\ }\href@noop {} {\textit{{\bibinfo {title} {Molecular {T}heory of
  {C}apillarity}}}}\ (\bibinfo  {publisher} {Oxford University Press},\ \bibinfo
  {address} {Oxford},\ \bibinfo {year} {1982})\BibitemShut {NoStop}%
\bibitem [{\citenamefont {Chapman}\ and\ \citenamefont
  {Cowling}(1970)}]{Chapman1970}%
  \BibitemOpen
  \bibfield  {author} {\bibinfo {author} {\bibfnamefont {S.}~\bibnamefont
  {Chapman}}\ and\ \bibinfo {author} {\bibfnamefont {T.}~\bibnamefont
  {Cowling}},\ }\href@noop {} {\textit{{\bibinfo {title} {The {M}athematical
  {T}heory of {N}on-{U}niform {G}ases}}}},\ \bibinfo {edition} {3rd}\ ed.\
  (\bibinfo  {publisher} {Cambridge University Press},\ \bibinfo {address}
  {Cambridge},\ \bibinfo {year} {1970})\BibitemShut {NoStop}%
\bibitem [{\citenamefont {Qian}\ \emph {et~al.}(1992)\citenamefont {Qian},
  \citenamefont {d'Humi\`{e}res},\ and\ \citenamefont {Lallemand}}]{Qian1992}%
  \BibitemOpen
  \bibfield  {author} {\bibinfo {author} {\bibfnamefont {Y.~H.}\ \bibnamefont
  {Qian}}, \bibinfo {author} {\bibfnamefont {D.}~\bibnamefont
  {d'Humi\`{e}res}}, \ and\ \bibinfo {author} {\bibfnamefont {P.}~\bibnamefont
  {Lallemand}},\ }\href@noop {} {\bibfield  {journal} {\bibinfo  {journal}
  {Europhys. Lett.}\ }\textbf {\bibinfo {volume} {17}},\ \bibinfo {pages} {479}
  (\bibinfo {year} {1992})}\BibitemShut {NoStop}%
\bibitem [{\citenamefont {Lallemand}\ and\ \citenamefont
  {Luo}(2000)}]{Lallemand2000}%
  \BibitemOpen
  \bibfield  {author} {\bibinfo {author} {\bibfnamefont {P.}~\bibnamefont
  {Lallemand}}\ and\ \bibinfo {author} {\bibfnamefont {L.-S.}\ \bibnamefont
  {Luo}},\ }\href@noop {} {\bibfield  {journal} {\bibinfo  {journal} {Phys.
  Rev. E}\ }\textbf {\bibinfo {volume} {61}},\ \bibinfo {pages} {6546}
  (\bibinfo {year} {2000})}\BibitemShut {NoStop}%
\bibitem [{\citenamefont {Huang}\ \emph
  {et~al.}(2019{\natexlab{a}})\citenamefont {Huang}, \citenamefont {Wu},\ and\
  \citenamefont {Adams}}]{Huang2019.STEOS}%
  \BibitemOpen
  \bibfield  {author} {\bibinfo {author} {\bibfnamefont {R.}~\bibnamefont
  {Huang}}, \bibinfo {author} {\bibfnamefont {H.}~\bibnamefont {Wu}}, \ and\
  \bibinfo {author} {\bibfnamefont {N.~A.}\ \bibnamefont {Adams}},\ }\href@noop
  {} {\bibfield  {journal} {\bibinfo  {journal} {Phys. Rev. E}\ }\textbf
  {\bibinfo {volume} {99}},\ \bibinfo {pages} {023303} (\bibinfo {year}
  {2019}{\natexlab{a}})}\BibitemShut {NoStop}%
\bibitem [{SM()}]{SM}%
  \BibitemOpen
  \href@noop {} {{\bibinfo {title} {See Supplemental Material 
   for technical details of the present mesoscopic LB model, which includes
  Refs.\ \cite{Huang2014, Huang2019.Grad}.}}}\BibitemShut {Stop}%
\bibitem [{\citenamefont {Huang}\ and\ \citenamefont {Wu}(2014)}]{Huang2014}%
  \BibitemOpen
  \bibfield  {author} {\bibinfo {author} {\bibfnamefont {R.}~\bibnamefont
  {Huang}}\ and\ \bibinfo {author} {\bibfnamefont {H.}~\bibnamefont {Wu}},\
  }\href@noop {} {\bibfield  {journal} {\bibinfo  {journal} {J. Comput. Phys.}\
  }\textbf {\bibinfo {volume} {274}},\ \bibinfo {pages} {50} (\bibinfo {year}
  {2014})}\BibitemShut {NoStop}%
\bibitem [{\citenamefont {Huang}\ \emph
  {et~al.}(2019{\natexlab{b}})\citenamefont {Huang}, \citenamefont {Wu},\ and\
  \citenamefont {Adams}}]{Huang2019.Grad}%
  \BibitemOpen
  \bibfield  {author} {\bibinfo {author} {\bibfnamefont {R.}~\bibnamefont
  {Huang}}, \bibinfo {author} {\bibfnamefont {H.}~\bibnamefont {Wu}}, \ and\
  \bibinfo {author} {\bibfnamefont {N.~A.}\ \bibnamefont {Adams}},\ }\href@noop
  {} {\bibfield  {journal} {\bibinfo  {journal} {Phys. Rev. E}\ }\textbf
  {\bibinfo {volume} {100}},\ \bibinfo {pages} {043306} (\bibinfo {year}
  {2019}{\natexlab{b}})}\BibitemShut {NoStop}%
\bibitem [{\citenamefont {Dellar}(2014)}]{Dellar2014}%
  \BibitemOpen
  \bibfield  {author} {\bibinfo {author} {\bibfnamefont {P.~J.}\ \bibnamefont
  {Dellar}},\ }\href@noop {} {\bibfield  {journal} {\bibinfo  {journal} {J.
  Comput. Phys.}\ }\textbf {\bibinfo {volume} {259}},\ \bibinfo {pages} {270}
  (\bibinfo {year} {2014})}\BibitemShut {NoStop}%
\bibitem [{\citenamefont {Geier}\ and\ \citenamefont
  {Pasquali}(2018)}]{Geier2018}%
  \BibitemOpen
  \bibfield  {author} {\bibinfo {author} {\bibfnamefont {M.}~\bibnamefont
  {Geier}}\ and\ \bibinfo {author} {\bibfnamefont {A.}~\bibnamefont
  {Pasquali}},\ }\href@noop {} {\bibfield  {journal} {\bibinfo  {journal}
  {Comput. Fluids}\ }\textbf {\bibinfo {volume} {166}},\ \bibinfo {pages} {139}
  (\bibinfo {year} {2018})}\BibitemShut {NoStop}%
\bibitem [{\citenamefont {Huang}\ \emph {et~al.}(2020)\citenamefont {Huang},
  \citenamefont {Lan},\ and\ \citenamefont {Li}}]{Huang2020}%
  \BibitemOpen
  \bibfield  {author} {\bibinfo {author} {\bibfnamefont {R.}~\bibnamefont
  {Huang}}, \bibinfo {author} {\bibfnamefont {L.}~\bibnamefont {Lan}}, \ and\
  \bibinfo {author} {\bibfnamefont {Q.}~\bibnamefont {Li}},\ }\href@noop {}
  {\bibfield  {journal} {\bibinfo  {journal} {Phys. Rev. E}\ }\textbf {\bibinfo
  {volume} {102}},\ \bibinfo {pages} {043304} (\bibinfo {year}
  {2020})}\BibitemShut {NoStop}%
\bibitem [{\citenamefont {Carnahan}\ and\ \citenamefont
  {Starling}(1969)}]{Carnahan1969}%
  \BibitemOpen
  \bibfield  {author} {\bibinfo {author} {\bibfnamefont {N.~F.}\ \bibnamefont
  {Carnahan}}\ and\ \bibinfo {author} {\bibfnamefont {K.~E.}\ \bibnamefont
  {Starling}},\ }\href@noop {} {\bibfield  {journal} {\bibinfo  {journal} {J.
  Chem. Phys.}\ }\textbf {\bibinfo {volume} {51}},\ \bibinfo {pages} {635}
  (\bibinfo {year} {1969})}\BibitemShut {NoStop}%
\bibitem [{\citenamefont {Huang}\ \emph
  {et~al.}(2019{\natexlab{c}})\citenamefont {Huang}, \citenamefont {Wu},\ and\
  \citenamefont {Adams}}]{Huang2019.TLBM}%
  \BibitemOpen
  \bibfield  {author} {\bibinfo {author} {\bibfnamefont {R.}~\bibnamefont
  {Huang}}, \bibinfo {author} {\bibfnamefont {H.}~\bibnamefont {Wu}}, \ and\
  \bibinfo {author} {\bibfnamefont {N.~A.}\ \bibnamefont {Adams}},\ }\href@noop
  {} {\bibfield  {journal} {\bibinfo  {journal} {J. Comput. Phys.}\ }\textbf
  {\bibinfo {volume} {392}},\ \bibinfo {pages} {227} (\bibinfo {year}
  {2019}{\natexlab{c}})}\BibitemShut {NoStop}%
\bibitem [{\citenamefont {Solomon}(1966)}]{Solomon1966}%
  \BibitemOpen
  \bibfield  {author} {\bibinfo {author} {\bibfnamefont {A.}~\bibnamefont
  {Solomon}},\ }\href@noop {} {\bibfield  {journal} {\bibinfo  {journal} {Math.
  Comp.}\ }\textbf {\bibinfo {volume} {20}},\ \bibinfo {pages} {347} (\bibinfo
  {year} {1966})}\BibitemShut {NoStop}%
\bibitem [{\citenamefont {Godsave}(1953)}]{Godsave1953}%
  \BibitemOpen
  \bibfield  {author} {\bibinfo {author} {\bibfnamefont {G.~A.~E.}\
  \bibnamefont {Godsave}},\ }\href@noop {} {\bibfield  {journal} {\bibinfo
  {journal} {Symp. (Int.) Combust.}\ }\textbf {\bibinfo {volume} {4}},\
  \bibinfo {pages} {818} (\bibinfo {year} {1953})}\BibitemShut {NoStop}%
\bibitem [{\citenamefont {Safari}\ \emph {et~al.}(2013)\citenamefont {Safari},
  \citenamefont {Rahimian},\ and\ \citenamefont {Krafczyk}}]{Safari2013}%
  \BibitemOpen
  \bibfield  {author} {\bibinfo {author} {\bibfnamefont {H.}~\bibnamefont
  {Safari}}, \bibinfo {author} {\bibfnamefont {M.~H.}\ \bibnamefont
  {Rahimian}}, \ and\ \bibinfo {author} {\bibfnamefont {M.}~\bibnamefont
  {Krafczyk}},\ }\href@noop {} {\bibfield  {journal} {\bibinfo  {journal}
  {Phys. Rev. E}\ }\textbf {\bibinfo {volume} {88}},\ \bibinfo {pages} {013304}
  (\bibinfo {year} {2013})}\BibitemShut {NoStop}%
\bibitem [{\citenamefont {Wen}\ \emph {et~al.}(2019)\citenamefont {Wen},
  \citenamefont {Kim}, \citenamefont {Shi}, \citenamefont {Wang}, \citenamefont
  {Man}, \citenamefont {Doi},\ and\ \citenamefont {Russell}}]{Wen2019}%
  \BibitemOpen
  \bibfield  {author} {\bibinfo {author} {\bibfnamefont {Y.}~\bibnamefont
  {Wen}}, \bibinfo {author} {\bibfnamefont {P.~Y.}\ \bibnamefont {Kim}},
  \bibinfo {author} {\bibfnamefont {S.}~\bibnamefont {Shi}}, \bibinfo {author}
  {\bibfnamefont {D.}~\bibnamefont {Wang}}, \bibinfo {author} {\bibfnamefont
  {X.}~\bibnamefont {Man}}, \bibinfo {author} {\bibfnamefont {M.}~\bibnamefont
  {Doi}}, \ and\ \bibinfo {author} {\bibfnamefont {T.~P.}\ \bibnamefont
  {Russell}},\ }\href@noop {} {\bibfield  {journal} {\bibinfo  {journal} {Soft
  Matter}\ }\textbf {\bibinfo {volume} {15}},\ \bibinfo {pages} {2135}
  (\bibinfo {year} {2019})}\BibitemShut {NoStop}%
\bibitem [{\citenamefont {Man}\ and\ \citenamefont {Doi}(2017)}]{Man2017}%
  \BibitemOpen
  \bibfield  {author} {\bibinfo {author} {\bibfnamefont {X.}~\bibnamefont
  {Man}}\ and\ \bibinfo {author} {\bibfnamefont {M.}~\bibnamefont {Doi}},\
  }\href@noop {} {\bibfield  {journal} {\bibinfo  {journal} {Phys. Rev. Lett.}\
  }\textbf {\bibinfo {volume} {119}},\ \bibinfo {pages} {044502} (\bibinfo
  {year} {2017})}\BibitemShut {NoStop}%
\bibitem [{\citenamefont {Fisher}\ \emph {et~al.}(1981)\citenamefont {Fisher},
  \citenamefont {Gamble},\ and\ \citenamefont {Middlehurst}}]{Fisher1981}%
  \BibitemOpen
  \bibfield  {author} {\bibinfo {author} {\bibfnamefont {L.~R.}\ \bibnamefont
  {Fisher}}, \bibinfo {author} {\bibfnamefont {R.~A.}\ \bibnamefont {Gamble}},
  \ and\ \bibinfo {author} {\bibfnamefont {J.}~\bibnamefont {Middlehurst}},\
  }\href@noop {} {\bibfield  {journal} {\bibinfo  {journal} {Nature}\ }\textbf
  {\bibinfo {volume} {290}},\ \bibinfo {pages} {575} (\bibinfo {year}
  {1981})}\BibitemShut {NoStop}%
\bibitem [{\citenamefont {Yang}\ \emph {et~al.}(2020)\citenamefont {Yang},
  \citenamefont {Sun}, \citenamefont {Fumagalli}, \citenamefont {Stebunov},
  \citenamefont {Haigh}, \citenamefont {Zhou}, \citenamefont {Grigorieva},
  \citenamefont {Wang},\ and\ \citenamefont {Geim}}]{Yang2020}%
  \BibitemOpen
  \bibfield  {author} {\bibinfo {author} {\bibfnamefont {Q.}~\bibnamefont
  {Yang}}, \bibinfo {author} {\bibfnamefont {P.~Z.}\ \bibnamefont {Sun}},
  \bibinfo {author} {\bibfnamefont {L.}~\bibnamefont {Fumagalli}}, \bibinfo
  {author} {\bibfnamefont {Y.~V.}\ \bibnamefont {Stebunov}}, \bibinfo {author}
  {\bibfnamefont {S.~J.}\ \bibnamefont {Haigh}}, \bibinfo {author}
  {\bibfnamefont {Z.~W.}\ \bibnamefont {Zhou}}, \bibinfo {author}
  {\bibfnamefont {I.~V.}\ \bibnamefont {Grigorieva}}, \bibinfo {author}
  {\bibfnamefont {F.~C.}\ \bibnamefont {Wang}}, \ and\ \bibinfo {author}
  {\bibfnamefont {A.~K.}\ \bibnamefont {Geim}},\ }\href@noop {} {\bibfield
  {journal} {\bibinfo  {journal} {Nature}\ }\textbf {\bibinfo {volume} {588}},\
  \bibinfo {pages} {250} (\bibinfo {year} {2020})}\BibitemShut {NoStop}%
\end{thebibliography}
%

\newpage


\section*{Supplementary material}

# Supplemental Material
# Mesoscopic Lattice Boltzmann Modeling of the Liquid-Vapor Phase Transition

Rongzong Huang,[1, *] Huiying Wu,[2, †] and Nikolaus A. Adams[3, ‡]

[1]*School of Energy Science and Engineering, Central South University, 410083 Changsha, China*
[2]*School of Mechanical Engineering, Shanghai Jiao Tong University, 200240 Shanghai, China*
[3]*Institute of Aerodynamics and Fluid Mechanics, Technical University of Munich, 85748 Garching, Germany*

(Dated: March 26, 2021)

In this Supplemental Material, the technical details of the present mesoscopic LB model for liquid-vapor phase transition are provided, of which the extensions to the three-dimensional lattice are straightforward. The discrete velocity for the two-dimensional nine-velocity (D2Q9) lattice is [1]

$$\mathbf{e}_i = \begin{cases} c(0,\ 0)^\mathrm{T}, & i = 0, \\ c\big(\cos[(i-1)\pi/2],\ \sin[(i-1)\pi/2]\big)^\mathrm{T}, & i = 1, 2, 3, 4, \\ \sqrt{2}c\big(\cos[(2i-1)\pi/4],\ \sin[(2i-1)\pi/4]\big)^\mathrm{T}, & i = 5, 6, 7, 8, \end{cases} \tag{1}$$

where $c = \delta_x/\delta_t$ is the lattice speed, with $\delta_x$ and $\delta_t$ the lattice spacing and time step, respectively. The corresponding orthogonal transformation matrix is [2]

$$\mathbf{M} = \begin{bmatrix} 1 & 1 & 1 & 1 & 1 & 1 & 1 & 1 & 1 \\ -4 & -1 & -1 & -1 & -1 & 2 & 2 & 2 & 2 \\ 4 & -2 & -2 & -2 & -2 & 1 & 1 & 1 & 1 \\ 0 & 1 & 0 & -1 & 0 & 1 & -1 & -1 & 1 \\ 0 & -2 & 0 & 2 & 0 & 1 & -1 & -1 & 1 \\ 0 & 0 & 1 & 0 & -1 & 1 & 1 & -1 & -1 \\ 0 & 0 & -2 & 0 & 2 & 1 & 1 & -1 & -1 \\ 0 & 1 & -1 & 1 & -1 & 0 & 0 & 0 & 0 \\ 0 & 0 & 0 & 0 & 0 & 1 & -1 & 1 & -1 \end{bmatrix}. \tag{2}$$

To recover a nonideal-gas EOS like $p_{\text{BE}}$, the equilibrium moment $\mathbf{m}^{\text{eq}}$ and discrete force $\mathbf{F}_m$ in the LB equation for density DF [i.e., Eq. (12b) in the main text] are devised as

$$\mathbf{m}^{\text{eq}} = [\rho,\ -2\rho + 2\eta + 3\rho|\hat{\mathbf{u}}|^2,\ \rho + \beta\eta - 3\rho|\hat{\mathbf{u}}|^2,\ \rho\hat{u}_x,\ -\rho\hat{u}_x,\ \rho\hat{u}_y,\ -\rho\hat{u}_y,\ \rho\hat{u}_x^2 - \rho\hat{u}_y^2,\ \rho\hat{u}_x\hat{u}_y]^\mathrm{T}, \tag{3a}$$

$$\mathbf{F}_m = [0,\ 6\hat{\mathbf{F}}\cdot\hat{\mathbf{u}},\ -6\hat{\mathbf{F}}\cdot\hat{\mathbf{u}},\ \hat{F}_x,\ -\hat{F}_x,\ \hat{F}_y,\ -\hat{F}_y,\ 2\hat{F}_x\hat{u}_x - 2\hat{F}_y\hat{u}_y,\ \hat{F}_x\hat{u}_y + \hat{F}_y\hat{u}_x]^\mathrm{T}, \tag{3b}$$

where $\hat{\mathbf{u}} = \mathbf{u}/c$, $\hat{\mathbf{F}} = \mathbf{F}/c$, $\beta = -2/(1-\varpi)$, and $\eta$ is a built-in variable. Here, $\varpi$ is a model coefficient related to the bulk viscosity. In the LB equation for total kinetic energy DF [i.e., Eq. (12c) in the main text], the equilibrium moment $\mathbf{n}^{\text{eq}}$ and discrete source $\mathbf{q}_m$ are devised as

$$\mathbf{n}^{\text{eq}} = [\rho e_k,\ -4\rho e_k + (4 + \gamma_1)C_{\text{ref}}T,\ 4\rho e_k - (4 - \gamma_2)C_{\text{ref}}T,\ \rho h_k \hat{u}_x,\ -\rho h_k \hat{u}_x,\ \rho h_k \hat{u}_y,\ -\rho h_k \hat{u}_y,\ 0,\ 0]^\mathrm{T}, \tag{4a}$$

$$\mathbf{q}_m = [q,\ \gamma_1 q,\ \gamma_2 q,\ q\hat{u}_x,\ -q\hat{u}_x,\ q\hat{u}_y,\ -q\hat{u}_y,\ 0,\ 0]^\mathrm{T}, \tag{4b}$$

where $C_{\text{ref}}$ is a reference *volumetric heat capacity*, $\rho h_k$ is named the *total kinetic enthalpy*, and $\gamma_{1,2}$ are model coefficients related to the heat conductivity. To correctly recover the meso-macroscopic governing equations derived from the kinetic model [i.e.,

* rongzong.huang@csu.edu.cn
† whysrj@sjtu.edu.cn
‡ nikolaus.adams@tum.de

Eq. (6) in the main text], the collision matrices $\mathbf{S}$ and $\mathbf{L}$ in the LB equations are set as [3, 4]

$$\mathbf{S} = \begin{bmatrix} s_0 & 0 & 0 & 0 & 0 & 0 & 0 & 0 & 0 \\ 0 & s_e & ks_\varepsilon \omega_e & 0 & h\hat{u}_x s_q \omega_e & 0 & h\hat{u}_y s_q \omega_e & 0 & 0 \\ 0 & 0 & s_\varepsilon & 0 & 0 & 0 & 0 & 0 & 0 \\ 0 & 0 & 0 & s_j & 0 & 0 & 0 & 0 & 0 \\ 0 & 0 & 0 & 0 & s_q & 0 & 0 & 0 & 0 \\ 0 & 0 & 0 & 0 & 0 & s_j & 0 & 0 & 0 \\ 0 & 0 & 0 & 0 & 0 & 0 & s_q & 0 & 0 \\ 0 & 0 & 0 & 0 & 2b\hat{u}_x s_q \omega_p & 0 & -2b\hat{u}_y s_q \omega_p & s_p & 0 \\ 0 & 0 & 0 & 0 & b\hat{u}_y s_q \omega_p & 0 & b\hat{u}_x s_q \omega_p & 0 & s_p \end{bmatrix}, \quad (5a)$$

$$\mathbf{L} = \begin{bmatrix} \sigma_0 & 0 & 0 & 0 & 0 & 0 & 0 & 0 & 0 \\ 0 & \sigma_e & 0 & 0 & 0 & 0 & 0 & 0 & 0 \\ 0 & 0 & \sigma_\varepsilon & 0 & 0 & 0 & 0 & 0 & 0 \\ 0 & 0 & 0 & \sigma_j & \sigma_q \omega_j & 0 & 0 & 0 & 0 \\ 0 & 0 & 0 & 0 & \sigma_q & 0 & 0 & 0 & 0 \\ 0 & 0 & 0 & 0 & 0 & \sigma_j & \sigma_q \omega_j & 0 & 0 \\ 0 & 0 & 0 & 0 & 0 & 0 & \sigma_q & 0 & 0 \\ 0 & 0 & 0 & 0 & 0 & 0 & 0 & \sigma_p & 0 \\ 0 & 0 & 0 & 0 & 0 & 0 & 0 & 0 & \sigma_p \end{bmatrix}, \quad (5b)$$

where $\omega_{e,p} = s_{e,p}/2 - 1$, $\omega_j = \sigma_j/2 - 1$, $k = 1 - \varpi$, $h = 6\varpi(1-\varpi)/(1-3\varpi)$, and $b = (1-\varpi)/(1-3\varpi)$. Furthermore, the relaxation parameters in $\mathbf{S}$ should satisfy $\tau_p \tau_q = (k+1)\tau_e \tau_q = 1/12$ with $\tau_{e,p,q} = s_{e,p,q}^{-1} - 1/2$. The last term in Eq. (12b) in the main text is introduced to compensate for the third-order discrete lattice effect with [5]

$$\mathbf{Q}_m = \frac{G^2 \delta_x^2 \delta_t^2}{12}[0,\ 6|\nabla\rho|^2,\ -6|\nabla\rho|^2,\ 0,\ 0,\ 0,\ 0,\ (\partial_x \rho)^2 - (\partial_y \rho)^2,\ \partial_x \rho\, \partial_y \rho]^T, \quad (6a)$$

and the last term in Eq. (12c) in the main text is introduced to account for the viscous dissipation with [6]

$$\mathbf{Y} = \begin{bmatrix} 0 & 0 & 0 & 0 & 0 & 0 & 0 & 0 & 0 \\ 0 & 0 & 0 & 0 & 0 & 0 & 0 & 0 & 0 \\ 0 & 0 & 0 & 0 & 0 & 0 & 0 & 0 & 0 \\ 0 & \hat{u}_x/3 & 0 & 0 & 0 & 0 & 0 & \hat{u}_x & 2\hat{u}_y \\ 0 & -\hat{u}_x/3 & 0 & 0 & 0 & 0 & 0 & -\hat{u}_x & -2\hat{u}_y \\ 0 & \hat{u}_y/3 & 0 & 0 & 0 & 0 & 0 & -\hat{u}_y & 2\hat{u}_x \\ 0 & -\hat{u}_y/3 & 0 & 0 & 0 & 0 & 0 & \hat{u}_y & -2\hat{u}_x \\ 0 & 0 & 0 & 0 & 0 & 0 & 0 & 0 & 0 \\ 0 & 0 & 0 & 0 & 0 & 0 & 0 & 0 & 0 \end{bmatrix}. \quad (6b)$$

Note that the $\nabla \rho$ in $\mathbf{Q}_m$ can be locally calculated by using $\mathbf{F}_{\text{pair}}$ [5], implying that the localizability of Eq. (12b) in the main text can be maintained.

[1] Y. H. Qian, D. d'Humières, and P. Lallemand, Europhys. Lett. **17**, 479 (1992).
[2] P. Lallemand and L.-S. Luo, Phys. Rev. E **61**, 6546 (2000).
[3] R. Huang, H. Wu, and N. A. Adams, Phys. Rev. E **99**, 023303 (2019).
[4] R. Huang and H. Wu, J. Comput. Phys. **274**, 50 (2014).
[5] R. Huang, H. Wu, and N. A. Adams, Phys. Rev. E **100**, 043306 (2019).
[6] R. Huang, H. Wu, and N. A. Adams, J. Comput. Phys. **392**, 227 (2019).



\end{document}